\documentclass[aps,pra]{revtex4}
\usepackage{epsfig,pstricks}
\usepackage{amsmath,amssymb}

\newtheorem{pdef}{Definition}[section]
\def\CC{{\rm\kern.24em \vrule width.04em height1.46ex depth-.07ex
\kern-.30em C}}
\def\RR{{\rm
         \vrule width.04em height1.58ex depth-.0ex
         \kern-.04em R}}
\def\id{{\rm 1\kern-.22em l}}

\newcommand{\beq}{\begin{equation}}
\newcommand{\beqa}{\begin{eqnarray}}
\newcommand{\nbeqa}{\begin{eqnarray*}}
\newcommand{\eeq}{\end{equation}}
\newcommand{\eeqa}{\end{eqnarray}}
\newcommand{\neeqa}{\end{eqnarray*}}
\newcommand{\bra}[1]{\left\langle #1 \right |}
\newcommand{\ket}[1]{\left | #1 \right\rangle}

\newcommand{\expect}[1]{\left\langle #1 \right\rangle}
\newcommand{\dexpect}[1]{\left\langle\hspace{-2mm}\left\langle\, 
		#1 \,\right\rangle\hspace{-2mm}\right\rangle}
\newcommand{\diag}{{\rm diag}\;}
\newcommand{\bigfrac}[2]{\mbox {${\displaystyle \frac{ #1 }{ #2 }}$}}

\begin{document}
\tighten

\title{Entanglement monotones and maximally entangled states 
 in multipartite qubit systems}
\author{Andreas Osterloh$^{1,2}$}
\author{Jens Siewert$^{1,3}$}
\affiliation{$^1$ MATIS-INFM $\&$ Dipartimento di Metodologie Fisiche e
    Chimiche (DMFCI), viale A. Doria 6, 95125 Catania, Italy\\
             $^2$ Institut f\"ur Theoretische Physik, 
                  Universit\"at Hannover, D-30167 Hannover, Germany\\
             $^3$ Institut f\"ur Theoretische Physik, 
                  Universit\"at Regensburg, D-93040 Regensburg, Germany}

\begin{abstract}
We present a method to construct entanglement measures 
for pure states of multipartite qubit systems. 
The key element of our approach
is an antilinear operator that we call {\em comb} in reference to the
{\em hairy-ball theorem}. 
For qubits (or spin 1/2) the combs are automatically 
invariant under $SL(2,\CC)$.
This implies that the {\em filters} obtained from the 
combs are entanglement monotones by construction.
We give alternative formulae for the concurrence and 
the 3-tangle as expectation values of certain antilinear operators.
As an application we discuss inequivalent
types of genuine four-, five- and six-qubit entanglement.

\end{abstract}

\keywords{Entanglement monotones; multipartite entanglement; antilinear operators}

\maketitle

\section{Introduction}

Entanglement is one the most striking features of quantum mechanics,
but it is also one of its most counterintuitive
consequences of which we still have rather incomplete knowledge~\cite{Bell87}.
Although the concentrated effort during the past decade has produced
an impressive progress, there is no general
qualitative and quantitative theory of entanglement.

A pure quantum-mechanical state of distinguishable particles is 
called disentangled with respect to a given 
partition ${\cal P}$ of the system {\em iff} it can be written as a 
tensor product of the parts of this partition. In the opposite case,
the state must contain some finite amount of entanglement. 
The question then is to characterize and quantify this entanglement.

As to {\em measuring} the amount of entanglement in a given pure
multipartite state, the first major step was made by
Bennett {\it et al.}~\cite{BennettDiVincenzo96} who discovered that the partial
entropy of a party in a bipartite quantum state is a measure of
entanglement. It coincides (asymptotically) with the entanglement of formation.
Subsequently, the entanglement of formation of a two-qubit state
was related to the concurrence~\cite{Hill97,Wootters98}.
Interestingly, by 
exploiting the knowledge of the mixed-state concurrence, the so-called
3-tangle $\tau_3$ which is a measure for three-partite pure states 
could be derived~\cite{Coffman00}.
This was a remarkable step since, loosely speaking, it opened the path 
to studying multipartite entanglement on solid grounds. 
Further, it was noticed by Uhlmann that antilinearity is
an important property of operators that measure 
entanglement~\cite{Uhlmann}.
A particularly interesting consequence of the 3-tangle formula
was presented by D\"ur {\it et al.} who found that there are two inequivalent
classes of sharing entanglement among three parties\cite{Duer00}.

Another important aspect of the research on entanglement measures 
was the question regarding the requirements for a function
that represents an entanglement monotone~\cite{MONOTONES}. It turned out
that the essential property to be satisfied is non-increasing behavior
on average under stochastic local operations and classical 
communication (SLOCC)~\cite{SLOCC,Duer00}.
Later, Verstraete {\em et al.} have demonstrated that all homogeneous positive
functions of pure-state density matrices that remain invariant
under determinant-one SLOCC operations are 
entanglement monotones~\cite{VerstraeteDM03}.

Despite the enormous effort, the only truly operational entanglement measure
for arbitrary mixed states at hand, up to now, is the concurrence.
For pure states we have a slightly farther view up to 
systems of two qutrits~\cite{Cereceda03,Briand03},
and for three qubits, due to the  3-tangle.
Various multipartite entanglement measures for pure-states have been proposed;
but most of these measures do not yield zero for all possible
product states (e.g.\ Refs.~\cite{Barnum03,HEYDARI04,Wallach,Wong00}).
This motivated the quest for an operational
entanglement measure emerging from one requirement only: that it be zero for
product states (not only for completely separable pure states).
In particular, the goal has been to explore the idea that entanglement
monotones are related to antilinear operators as pointed out for the
concurrence by Uhlmann~\cite{Uhlmann}.
Here we show that it is possible to construct a {\em filter},
i.e., an operator that has zero expectation value for all product states. 
It will turn out that these filters are entanglement monotones 
by construction. Interestingly, the two-qubit concurrence
and the 3-tangle have various equivalent filter representations (see below).
In order to illustrate the application of the method to a
nontrivial example, we will present filters for
up to six-qubit states that are able to distinguish inequivalent 
types of genuine multipartite entanglement. 

Before finding a measure for genuine multipartite entanglement,
one first has to agree about a definition of maximal
multipartite entanglement: 
\begin{pdef}\label{def}
A pure $q$-qubit state $\ket{\psi_q}$ 
has {\em maximal genuine multipartite entanglement},
i.e. {\em q-tangle}, if and only if
\begin{itemize} 
\item[(i)]{All reduced density matrices of $\ket{\psi_q}$ with rank
      $\leq 2$
      (this includes all $(q-1)$-site and single-site ones) 
      are maximally mixed}
\item[(ii)]{all $p$-site reduced density matrices of $\ket{\psi_q}$,
      have zero $p$-tangle; $1<p<q$.}
\item[(iii)]{there is a canonical form of any maximally $q$-tangled state, 
      for which properties (i) and (ii) are unaffected by phase factors, i.e.
      they are phase invariant.}
\end{itemize}
\end{pdef}
A stronger form of condition (i) appeared in Ref.~\cite{Gisin98}, 
where it is demanded that all reduced density matrices be maximally mixed.

Notice that the first 
condition induces that all reduced density matrices of the state have 
rank larger than $1$. This excludes product states of whatsoever kind.
We emphasize that we use the term 
{\em genuine $q$-qubit entanglement} in a more restricted sense
than, e.g., in Ref.~\cite{Duer00}; in particular, the only class of 
three-qubit states with genuine three-partite entanglement is represented
by the GHZ state.

Some remarks are in order: whereas the first two requirements are well motivated,
since the first means a maximal gain of information when a bit of information
is read out of a maximally entangled state, and the second excludes hybrids
of many different types of entanglement (somewhat following the idea of
entanglement as a resource whose amount can be distributed among 
possibly different types of entanglement only; see e.g. Ref.~\cite{Coffman00}),
we have no good argument in favor of the third, except that maximally 
entangled states for two and three qubits have such a canonical form.  

\section{Combs and filters}

The basic concept is that of the {\em comb}.
We define a {\em comb of first order} 
as an antilinear operator $A$ with 
zero expectation value for all states of a certain Hilbert space
${\cal H}$. 
That is,
\begin{equation}\label{def:Toeter}
\bra{\psi} A \ket{\psi}=\bra{\psi} L C\ket{\psi}
	=\bra{\psi} L \ket{\psi^*}\equiv 0
\end{equation}
for all $\ket{\psi}\in{\cal H}$, where $L$ is a linear operator and 
$C$ is the complex conjugation.
Here $A$ necessarily has to be antilinear 
(a linear operator with this property is zero itself).
For simplicity we abbreviate
\begin{equation}
\bra{\psi} L C\ket{\psi}=:\expect{L}_C \ \ .
\label{defexpect}
\end{equation}
Note that the complex conjugation is {\em included}
in the definition of the expectation value $\langle \ldots \rangle_C$
in Eq.\ (\ref{defexpect}).

We will use the comb operators~\footnote{
                If the antilinear operator $A=L C$ is 
                a comb (with the complex conjugation $C$),
                for the sake of brevity
                we will also call the linear operator $L$
                a comb.}
in order to construct the desired {\em filters} 
which are defined as antilinear
operators whose expectation values vanish for all product states.
While a comb is a local, i.e., a single-qubit operator, 
a filter is a non-local operator that acts 
on the whole multi-qubit state.
It is worth mentioning already at this point 
that such a filter is invariant under 
${\cal P}$-local unitary 
transformations if the combs have this property. 
Even more, it is invariant under the complex 
extension of the corresponding unitary group 
which is isomorphic to the special linear group. 
Since the latter represents the 
SLOCC operations for qubits~\cite{SLOCC,Duer00},
the filters will be entanglement monotones by construction.

We focus on multipartite systems of
qubits (i.e., spin 1/2). The local Hilbert space is 
${\cal H}_j=\CC^2=:{\frak h}$ for all $j$. We need the Pauli matrices
$\sigma_0:=\id$, $\sigma_1:=\sigma_x$,
$\sigma_2:=\sigma_y$, and
$\sigma_3:=\sigma_z$.
It is straight forward to verify that the only 
single-qubit comb is the operator
$\sigma_y$:
\[
\bra{\psi} \sigma_y C\ket{\psi}=\expect{\sigma_y}_C\equiv 0 \ \ .
\]
Since its expectation value is a bi(anti-)linear expression in the
coefficients of the state we denote it a comb
of {\em order 1}. In general we will call a comb to
be {\em of order n} if its expectation value is $2n$-linear
in the coefficients of the state.
There is one independent single-qubit comb which is of 2nd order. 
One can verify that for an arbitrary single-qubit state
\begin{equation}
0=\expect{\sigma_\mu}_C\expect{\sigma^\mu}_C:=
\sum_{\mu,\nu=0}^3 
\expect{\sigma_\mu}_C
		g^{\mu,\nu}\expect{\sigma_\nu}_C\; ,
\end{equation}
with
$g^{\mu,\nu}=\diag\{-1,1,0,1\}$
being very similar to the Minkowski metric.\footnote{Note that the only $SL(2,\CC)$
invariant {\bf linear} operator of order two is very similar to
its antilinear counterpart: $0=\expect{\sigma_\mu}\expect{\sigma^\mu}:=
\sum_{\mu,\nu=0}^3 
\expect{\sigma_\mu}
		g^{\mu,\nu}\expect{\sigma_\nu}$ with 
$g^{\mu,\nu}=\diag\{-1,1,1,1\}$.}
Both combs are $SL(2,\CC)$ invariant~\cite{OScombs}.

It will prove useful to introduce the embedding
\begin{equation}\label{embedding}
{\cal E}_n\; :\; 
\begin{array}{ccc}
{\cal H}   &\hookrightarrow& \mathfrak{H}_n={\cal H}^{\otimes\; n} \\
\ket{\psi} &\longrightarrow& {\cal E}_n\ket{\psi} = \ket{\psi}^{\otimes\; n}
\ \ .
\end{array}
\end{equation}
Further define the product $\bullet$ for operators 
$O$, $P$: ${\cal H} \longrightarrow {\cal H}$ such that
\begin{equation}\label{embedding:product}
O \bullet P\; :\; 
\begin{array}{ccc}
\mathfrak{H}_2   &\rightarrow& \mathfrak{H}_2 \\
O \bullet P {\cal E}_2(\ket{\psi}) &=& O\ket{\psi} \otimes P\ket{\psi}
\ \ .
\end{array}
\end{equation}
Then we have the single-site (${\cal H}=\CC^2$) 
comb $\sigma_y$ for $\mathfrak{H}_1={\cal H}$
and $\sigma_\mu\bullet\sigma^\mu$ for $\mathfrak{H}_2$.

These two one-site combs are sufficient to construct
filters for multipartite qubit systems, which are entanglement monotones 
by construction. For $n$-qubit filters we will 
use the symbol ${\cal F}^{(n)}$. 
Filters for two qubits are
\begin{eqnarray}\label{2-filters}
{\cal F}^{(2)}_1\ &=&\ \sigma_y\otimes \sigma_y \\
{\cal F}^{(2)}_2\ &=&\ \frac{1}{3}\ (\sigma_\mu\otimes\sigma_\nu)\bullet
		  (\sigma^\mu\otimes \sigma^\nu)\ \ .
\end{eqnarray}
Both forms are explicitly permutation invariant, and they are filters
since, if the state were a product, the combs would annihilate
its expectation value.
From the filters we obtain the pure-state concurrence in two different
equivalent forms:
\begin{eqnarray}\label{2-measures}
C\ &=&\ \left |\dexpect{{\cal F}^{(2)}_1}_C\right| \\
{\hspace*{-2mm}C}^2\ &=&\  \left|\dexpect{{\cal F}^{(2)}_2}_C\right|
\ \equiv \ \frac{1}{3}\ \left |\expect{\sigma_\mu\otimes\sigma_\nu}_C
    \expect{\sigma^\mu\otimes\sigma^\nu}_C\right| \ \ . \nonumber
\end{eqnarray}
While the first form in Eq.\ (\ref{2-measures}) 
has the well-know
convex-roof extension of the pure-state concurrence via the 
matrix~\cite{Hill97,Wootters98,Uhlmann} 
\begin{equation}\label{mixed1}
R \ = \ \sqrt{\rho}\ \sigma_y\otimes\sigma_y\  
                           \rho^*\  \sigma_y\otimes\sigma_y \
    \sqrt{\rho}
\end{equation}
it can be shown that the convex roof extension of the second form 
in Eq.\ (\ref{2-measures}) is related to 
\begin{eqnarray}\label{mixed2}
   Q &=& \sqrt{\rho}\ \sigma_\mu\otimes\sigma_\nu \ \rho^*\  
                        \sigma_\kappa\otimes\sigma_\lambda\\
        &&\quad  \rho\ \sigma^\mu\otimes\sigma^\nu\  \rho^*\  
                       \sigma^\kappa\otimes\sigma^\lambda
          \ \sqrt{\rho} \ \ .\nonumber
\end{eqnarray}
and we find that $Q\equiv R^2$.

Now let us consider the 3-tangle~\cite{Coffman00}.
For states of three qubits we find, e.g.,
\begin{eqnarray}\label{3-filters}
{\cal F}^{(3)}_1 &\ =&\ (\sigma_\mu\otimes\sigma_y\otimes\sigma_y)\bullet
		  (\sigma^\mu\otimes \sigma_y\otimes\sigma_y)\\
{\cal F}^{(3)}_2 &\ =&\ \bigfrac{1}{3}\
		(\sigma_\mu\otimes\sigma_\nu\otimes\sigma_\lambda)\bullet
		  (\sigma^\mu\otimes \sigma^\nu\otimes\sigma^\lambda)
 \ \ . \ \ 
\end{eqnarray}
Both ${\cal F}^{(3)}_1$ and ${\cal F}^{(3)}_2$
are filters and the latter is explicitly permutation invariant. 
From these operators the pure-state 3-tangle is obtained in the following way:
\begin{eqnarray}\label{3-measures}
\tau_3&=&\left |\dexpect{{\cal F}^{(3)}_1}_C\right| 
      = \left|\dexpect{{\cal F}^{(3)}_2}_C\right|
\end{eqnarray}
Interestingly, {\em all} three-qubit
filters are powers of the 3-tangle as entanglement measure.
We mention, however, that there is no immediate extension to mixed states 
as in the case of the 'alternative' two-qubit concurrence, 
\mbox{Eq.\ (\ref{mixed2})}.

\section{Filters for four-qubit states}
Classifications of four-qubit states with respect
to their entanglement properties have been studied, e.g., in 
Refs.~\cite{VerstraeteDMV02,BriandLT03,Miyake03}. Here we introduce
three four-qubit filter operators and study the three classes of entangled
states they are measuring.

A four-qubit filter has the property that its expectation value
for a given state is zero if the state is separable, i.e.,
if there is a one-qubit or a two-qubit part which can be factored out
(note that for a three-qubit filter it is enough to extract
one-qubit parts only). An expression that obeys this requirement
for any single qubit and any combination of qubit pairs is given by
\begin{equation}
\label{fourbit6lin}
{\cal F}^{(4)}_1  = 
                (\sigma_\mu\sigma_\nu\sigma_y\sigma_y)\bullet
                  (\sigma^\mu\sigma_y\sigma_\lambda\sigma_y)
        \bullet(\sigma_y\sigma^\nu\sigma^\lambda\sigma_y) \ \ .
\end{equation}
Recall that any combination of the type 
$\sigma_{\mu}\sigma_y$ ($\mu\ne 2$) 
represents a two-qubit comb.
Note that the expectation value of an $n$th-order four-qubit 
     filter has to be taken with
     respect to the corresponding ${\mathfrak H}_n$, 
     see Ref.~\cite{OScombs}). 
It is straightforward to check that for a four-qubit GHZ state
\begin{equation}
\label{ghz4}
    \ket{\Phi_1}\ =\ \frac{1}{\sqrt{2}}( \ket{0000}  +  \ket{1111})
\end{equation}
we have  $\bra{\Phi_1}{\cal F}^{(4)}_1 \ket{\Phi_1^*}=1$. 
However, there is another state for which
$\langle{\cal F}^{(4)}_1 \rangle_C$ does not vanish. For
\begin{equation}
\label{wlength6}
    \ket{\Phi_5}\ =\ \frac{1}{\sqrt{6}}
    (\sqrt{2}\ket{1111}+\ket{1000}+\ket{0100}+\ket{0010}+\ket{0001})
\end{equation}
we find  $\bra{\Phi_5}{\cal F}^{(4)}_1 \ket{\Phi_5^*}=8/9$.
Interestingly, (\ref{wlength6}) is the only maximally entangled state 
measured by the four qubit hyperdeterminant\cite{Miyake03}, which is a 
homogeneous function of degree $24$. 
Its value for this state is $(\bigfrac{8}{9})^4$, i.e. exactly the same
as of the 24th order homogeneous invariant $\left({\cal F}^{(4)}_1\right)^4$,
which however also measures the GHZ state. The hyperdeterminant of
the four qubit GHZ state is zero.

Besides the 3rd-order filter ${\cal F}^{(4)}_1$ there exist also
filters of 4th order  and of 6th order. Examples are
\begin{eqnarray}\label{4-filters}
{\cal F}^{(4)}_2 &=&
                (\sigma_\mu\sigma_\nu\sigma_y\sigma_y)\bullet
                  (\sigma^\mu\sigma_y\sigma_\lambda\sigma_y)\bullet
   \nonumber\\
   &&\qquad\qquad \bullet        (\sigma_y\sigma^\nu\sigma_y\sigma_\tau) 
            \bullet
                (\sigma_y\sigma_y\sigma^\lambda\sigma^\tau)\\
{\cal F}^{(4)}_3 &=&\bigfrac{1}{2}
                (\sigma_\mu\sigma_\nu\sigma_y\sigma_y)\bullet
                  (\sigma^\mu\sigma^\nu\sigma_y\sigma_y)
\bullet(\sigma_\rho\sigma_y\sigma_\tau\sigma_y) \bullet
         \nonumber\\ &&\qquad  \bullet    
          (\sigma^\rho\sigma_y\sigma^\tau\sigma_y)
 \bullet(\sigma_y\sigma_\rho\sigma_\tau\sigma_y) \bullet
                (\sigma_y\sigma^\rho\sigma^\tau\sigma_y)
 \ \ . \nonumber
\end{eqnarray}
While ${\cal F}^{(4)}_2$ measures only GHZ-type entanglement
($\bra{\Phi_2}{\cal F}^{(4)}_2 \ket{\Phi_2^*}=1$)
the 6th-order filter ${\cal F}^{(4)}_3$ has the non-zero expectation values 
1/2 for the GHZ state and 1 for yet another state,
\begin{equation}
\label{wlength4}
    \ket{\Phi_4}\ =\ \frac{1}{2}
    (\ket{1111}+\ket{1100}+\ket{0010}+\ket{0001})\ \ .
\end{equation}
${\cal F}^{(4)}_1$ and ${\cal F}^{(4)}_2$ have zero expectation value 
for this state (as well as the hyperdeterminant).
Finally, all four-qubit filters ${\cal F}_j^{(4)}$ ($j=1,2,3$) 
have zero expectation value for the four-qubit {\em W} state 
$1/2(\ket{0111}+\ket{1011}+\ket{1101}+\ket{1110})$.

The states $\ket{\Phi_j}$ are the maximally entangled states
for four qubits; they satisfy all three requirements in Def. \ref{def},
including the stronger condition (i) from Ref.~\cite{Gisin98}.
Note that they cannot be transformed into one another by SLOCC operations:
A state with a finite expectation value for one filter cannot be
transformed by means of SLOCC operations into a state with zero expectation
value for the same filter.
For example, ${\cal F}^{(4)}_2$ detects the GHZ state $\ket{\Phi_1}$
but gives zero for the other two states. Therefore, the four-qubit
entanglement in those states must be different from that of the GHZ state.

Hence, there are at least three inequivalent types of genuine entanglement
for four qubits~\footnote{In fact, there are {\em exactly} three 
                         maximally entangled states for four qubits. 
                         This will be discussed in a forthcoming publication.}.
We mention that the three maximally entangled states $\ket{\Phi_j}$
are not distinguished by the classification for pure four-qubit states
of Ref.~\cite{VerstraeteDMV02}. This can be seen by computing 
the expectation values of the four-qubit filters and the
reduced one-qubit density matrices for each of the nine class representatives
of Ref.~\cite{VerstraeteDMV02}. Only the classes 1--4 and 6 have 
non-vanishing ``4-tangle''. The corresponding local density matrices
can be completely mixed {\em only} for class 1. Therefore, all three
states $\ket{\Phi_j}$ must belong to that class. 

\section{Filters for more qubits}

In this section we will continue the discussion from the previous
section and demonstrate how general multipartite filters are constructed.
It is not the scope of this work
to discuss independence and completeness of a given set of filters,
nor to ``taylor'' a filter for a given single class of entanglement.
We only emphasize that every filter is an invariant
and that linear homogeneous combinations and in fact any homogeneous
function of them is an invariant, as well.
Thus, when a sufficient set of independent filters
is known together with their weights for the corresponding entanglement 
classes, such a taylored invariant can be constructed.
This invariant, though, is not expected to be simply
the modulus square of some filter.

For five qubits we find four independent filters, their independence becoming 
clear from their values on a set of maximally entangled states.
In order to compactify the formulas,
the tensor product symbol $\otimes$ will be omitted.

\beqa\label{5-filters}
{\cal F}^{(5)}_1 &=&(\sigma_{\mu_1}\sigma_{\mu_2}\sigma_{\mu_3}\sigma_y\sigma_y)\bullet
		  (\sigma^{\mu_1}\sigma^{\mu_2}\sigma_y\sigma_{\mu_4}\sigma_y)\\
&& \qquad \bullet (\sigma_{\mu_5}\sigma_y\sigma^{\mu_3}\sigma^{\mu_4}\sigma_y)
 \bullet (\sigma^{\mu_5}\sigma_y\sigma_y\sigma_y\sigma_y)\nonumber\\
{\cal F}^{(5)}_2 &=&(\sigma_{\mu_1}\sigma_{\mu_2}\sigma_{\mu_3}\sigma_y\sigma_y)\bullet
		  (\sigma^{\mu_1}\sigma_y\sigma_y\sigma_{\mu_4}\sigma_{\mu_5})\\
&& \qquad \bullet(\sigma_y\sigma^{\mu_2}\sigma_y\sigma_y\sigma_y)
 \bullet(\sigma_y\sigma_y\sigma^{\mu_3}\sigma_y\sigma_y) \nonumber \\
&& \qquad \bullet (\sigma_y\sigma_y\sigma_y\sigma^{\mu_4}\sigma_y)
 \bullet (\sigma_y\sigma_y\sigma_y\sigma_y\sigma^{\mu_5})\nonumber \\
{\cal F}^{(5)}_3 &=&(\sigma_{\mu_1}\sigma_{\mu_2}\sigma_{\mu_3}\sigma_y\sigma_y)\bullet
		  (\sigma^{\mu_1}\sigma^{\mu_2}\sigma_{\mu_4}\sigma_y\sigma_y)\\
&& \qquad \bullet (\sigma_{\mu_5}\sigma_y\sigma^{\mu_3}\sigma_{\mu_6}\sigma_y)\bullet
		  (\sigma^{\mu_5}\sigma_y\sigma^{\mu_4}\sigma_{\mu_7}\sigma_y)\nonumber\\
&& \qquad \bullet (\sigma_{\mu_8}\sigma_y\sigma_y\sigma^{\mu_6}\sigma_{\mu_9})\bullet
		  (\sigma^{\mu_8}\sigma_y\sigma_y\sigma^{\mu_7}\sigma^{\mu_9})\nonumber\\
{\cal F}^{(5)}_4 &=&\bigfrac{1}{8}(\sigma_{\mu_1}\sigma_{\mu_2}\sigma_{\mu_3}\sigma_y\sigma_y)\bullet
		  (\sigma^{\mu_1}\sigma^{\mu_2}\sigma^{\mu_3}\sigma_y\sigma_y)\\
&& \qquad \bullet (\sigma_{\mu_4}\sigma_y\sigma_{\mu_5}\sigma_{\mu_6}\sigma_y)\bullet
		  (\sigma^{\mu_4}\sigma_y\sigma^{\mu_5}\sigma^{\mu_6}\sigma_y)\nonumber\\
&& \qquad \bullet (\sigma_{\mu_7}\sigma_y\sigma_y\sigma_{\mu_8}\sigma_{\mu_9})\bullet
		  (\sigma^{\mu_7}\sigma_y\sigma_y\sigma^{\mu_8}\sigma^{\mu_9})\nonumber
\eeqa

The set of maximally entangled states distinguished by these filters is

\beqa
\label{w5length2}
    \ket{\Psi_2}&=&\frac{1}{\sqrt{2}}
    (\ket{11111}+\ket{00000})\\
\label{w5length4}
    \ket{\Psi_4}&=&\frac{1}{2}
    (\ket{11111}+\ket{11100}+\ket{00010}+\ket{00001})\\
\label{w5length5}
    \ket{\Psi_5}&=&\frac{1}{\sqrt{6}}
    (\sqrt{2}\ket{11111}+\ket{11000}+\ket{00100}+\ket{00010}+\ket{00001})\\
\label{w5length6}
    \ket{\Psi_6}&=&\frac{1}{2\sqrt{2}}
    (\sqrt{3}\ket{11111}+\ket{10000}+\ket{01000}+\ket{00100}+\ket{00010}+\ket{00001})
\eeqa
The index of the state indicates the number of Fock-states in the normal form
of the state and will be termed its {\em length}; 
a deeper discussion of the maximally entangled states and their connection to
the filters that measure them is beyond the scope of this article and will
be reported elsewhere.\\
The states $\ket{\Psi_2}$ -- $\ket{\Psi_5}$ satisfy all three
requirements of definition \ref{def} for being a maximally entangled states.
It is interesting that $\ket{\Psi_6}$ instead satisfies only the 1st and the 3rd requirement
but contains fourtangle as measured by the filter ${\cal F}^{(4)}_3$.

For six quibits we only exemplarily write two independent filters
but indicate how to construct filters for a general number of qubits.

\beqa\label{6-filters}
{\cal F}^{(6)}_1 &=&(\sigma_{\mu_1}\sigma_{\mu_2}\sigma_y\sigma_y\sigma_y\sigma_y)\bullet
		  (\sigma^{\mu_1}\sigma_y\sigma_{\mu_3}\sigma_y\sigma_y\sigma_y)\\
&& \qquad \bullet(\sigma_{\mu_6}\sigma_y\sigma_y\sigma_{\mu_4}\sigma_y\sigma_y)
 \bullet(\sigma_y\sigma_y\sigma^{\mu_3}\sigma_y\sigma_{\mu_5}\sigma_y) \nonumber \\
&& \qquad \bullet (\sigma^{\mu_6}\sigma^{\mu_2}\sigma_y\sigma^{\mu_4}\sigma^{\mu_5}\sigma_y)\nonumber
\\
{\cal F}^{(6)}_2 &=&(\sigma_{\mu_1}\sigma_{\mu_2}\sigma_y\sigma_y\sigma_y\sigma_y)\bullet
		  (\sigma^{\mu_1}\sigma_y\sigma_{\mu_3}\sigma_y\sigma_y\sigma_y)\\
&& \qquad \bullet(\sigma_{\mu_6}\sigma^{\mu_2}\sigma^{\mu_3}\sigma_{\mu_4}\sigma_y\sigma_y)
 \bullet(\sigma_y\sigma_y\sigma_y\sigma^{\mu_4}\sigma_{\mu_5}\sigma_y) \nonumber \\
&& \qquad \bullet (\sigma^{\mu_6}\sigma_y\sigma_y\sigma_y\sigma^{\mu_5}\sigma_y)\nonumber
\\
&\vdots& \nonumber \\
{\cal F}^{(6)}_i &=&(\sigma_{\mu_\bullet}\sigma_{\mu_\bullet}\sigma_y\sigma_y\sigma_y\sigma_y)\bullet
		  (\sigma_{\mu_\bullet}\sigma_y\sigma_{\mu_\bullet}\sigma_y\sigma_y\sigma_y)\\
&& \qquad \bullet(\sigma_{\mu_\bullet}\sigma_\bullet \sigma_\bullet \sigma_{\mu_\bullet}\sigma_y\sigma_y)
 \bullet(\sigma_{\mu_\bullet}\sigma_\bullet \sigma_\bullet \sigma_\bullet 
\sigma_{\mu_\bullet}\sigma_y)\bullet (\sigma_\bullet \sigma_\bullet \sigma_\bullet 
\sigma_\bullet \sigma_\bullet \sigma_\bullet)\dots \nonumber 
\eeqa
where in the latter formula all the $\mu_\bullet$ are to be
contracted properly; in the $\sigma_\bullet$ the ``$\bullet$'' either 
have to be substituted by indices which then have to be contracted properly or 
by $\sigma_y$. 
This also indicates how higher filters can be constructed
and suggests that for a filter of an $n$-qubit system, at least
$\mathfrak{H}_{n-1}$ be needed. 
It is worthwhile to mention that the above list is not
meant to be exhaustive, nor did we explicitly check for permutation invariance,
which eventually could help cristalizing the ``proper'' filters.
The set of maximally entangled states to be distinguished by the six 
qubit filters is
\beqa
\label{w6length2}
    \ket{\Xi_2}&=&\frac{1}{\sqrt{2}}
    (\ket{111111}+\ket{000000})\\
\label{w6length4}
    \ket{\Xi_4}&=&\frac{1}{2}
    (\ket{111111}+\ket{111100}+\ket{000010}+\ket{000001})\\
\label{w6length5}
    \ket{\Xi_5}&=&\frac{1}{\sqrt{6}}
    (\sqrt{2}\ket{111111}+\ket{111000}+\ket{000100}+\ket{000010}+\ket{000001})\\
\label{w6length6}
    \ket{\Xi_6}&=&\frac{1}{2\sqrt{2}}
    (\sqrt{3}\ket{1\dots 1}+\ket{110000}+\ket{00}\otimes\ket{W_4})\\
\label{w6length7}
    \ket{\Xi_7}&=&\frac{1}{2\sqrt{2}}
    (\sqrt{3}\ket{111111}+\ket{W_6})
\eeqa
where $\ket{W_4}:=\ket{1000}+\ket{0100}+\ket{0010}+\ket{0001}$ and
$\ket{W_6}$ analoguously
are the W state for four and six qubits.
We want to mention that only the states up to length $5$ are free of any 
subtangle and that only for states up to length $4$ all reduced
density matrices are maximally mixed~\cite{Gisin98}.

The filter values for the maximally entangled states
are reported in the table. The states are classified by the
length of their normal form. An ``X'' indicates that the 
corresponding state does not occur.
Whereas the tangles for
\begin{table}[ph]
\label{table}
\begin{center}
\begin{tabular}{@{}|c||c|c|c||c|c|c|c||c|c|@{}}
\hline
length & $|{\cal F}^{(4)}_1|$ & $|{\cal F}^{(4)}_2|$ & 
$|{\cal F}^{(4)}_3|$ & $|{\cal F}^{(5)}_1|$ & $|{\cal F}^{(5)}_{2}|$ 
& $|{\cal F}^{(5)}_3|$ & $|{\cal F}^{(5)}_4|$ & $|{\cal F}^{(6)}_1|$ & $|{\cal F}^{(6)}_2|$ \\
\hline\hline
2 & 1               & 1& $\bigfrac{1}{2}$& 1                        & 1& 1                    & 
            $\bigfrac{1}{8}$ & 1 & 1\\
4 & 0               & 0& 1               & 0                        & 0& 0                    & 
                           1 & 0 & 0\\
5 & $\bigfrac{8}{9}$& 0& 0               & 0                        & 0& $\bigfrac{2^6}{3^5}$ & 
                           0 & 0 & 0 \\
6 &  X              & X& X               & $\bigfrac{3\sqrt{3}}{32}$& 0& 0                    & 
                           0 & 0 & 0\\
7 &  X              & X& X               & X                        & X& X                    & 
                           X & 0 &$\bigfrac{2^8}{5^5}$\\
\hline
\end{tabular} 
\end{center}
\end{table}
four/five qubits discriminate all three/four maximally entangled states, 
the two six-tangles we explicitely wrote only attribute to those
states with minimal length (the GHZ) and with maximal length. 
This table shows that the indicated states correspond to different
entanglement SLOCC classes.
In fact there is a relation between the length of the state and the 
degree of multilinearity of the filter, which will be reported on in another 
publication.

\section{Conclusions}
We have presented a new and efficient way
of generating entanglement monotones. It is based on
operators which we called {\em filters}.
The expectation values of these operators are zero for
all possible  product states, not only for the completely factoring case.
The building blocks of the filters (denoted
{\em combs}) guarantee invariance under $SL(2,\CC)^{\otimes N}$ for qubits. As
a consequence, all filters are automatically entanglement monotones.
They are measures of genuine multipartite entanglement.
This circumvents the difficult task to construct entanglement 
monotones from the essentially known (linear) local unitary invariants.

As an immediate result of our method the concurrence 
for pure two-qubit states is reproduced.
Moreover, we have found an alternative expression for 
the concurrence with the corresponding convex roof extension
based on the corresponding filter operator.
The application of the method to pure three-qubit states
yields several operator-based expressions for the 3-tangle,
including an explicitly permutation-invariant form.

Further advantages of this approach are the feasibility of constructing
specific monotones that vanish for certain separable (pure) states
and the applicability of this concept to partitions into subsystems 
other than qubits (i.e. qutrits\ldots).
The methods permits in a direct manner quantification and
classification of multipartite entanglement.
We demonstrate this with the explicit expressions for four- up to six-qubit 
entanglement measures that for the first time detect three different types 
of genuine four-qubit entanglement and four different types of five-qubit 
entanglement; the types of genuine four-qubit entanglement are not distinguished 
by the classification of four-qubit states in Ref.~\cite{VerstraeteDMV02}.

As to $N$-qubit systems, there remain various interesting questions.
Clearly, it would be desirable to have a recipe how to 
build invariant combs for more complicated systems (e.g. higher spin).
It would also be interesting to know what characterizes
a complete set of filters for any given $N$.
While it is not obvious how the convex roof construction for two qubits
can be generalized, we believe that the operator form of the
$N$-tangles in terms of filters makes it easier to solve this problem.
The question is whether there is a systematic way to obtain a
convex-roof construction for a given filter with general
multi-linearity.

{\em Acknowledgments -- }
We would like to thank L.\ Amico, R.\ Fazio, and especially A.\ Uhlmann
for stimulating discussions. This work was supported by the EU 
RTN grant HPRN-CT-2000-00144, the Vigoni Program of the German
Academic Exchange Service and the Sonderforschungsbereich 631 
of the German Research Foundation. J.S. holds a Heisenberg
fellowship from the German Research Foundation.

%\bibliography{measures}

\end{document}